\documentclass[aip,rsi,amsmath,amssymb,reprint]{revtex4-1}
\usepackage{graphicx, color}

\begin{document}
\title{On-chip terahertz modulation and emission with integrated graphene junctions}

\author{Joshua O. Island}
\email{jisland@physics.unlv.edu}
\affiliation{Department of Physics, University of California, Santa Barbara, CA, 93106 USA}
\affiliation{Department of Physics and Astronomy, University of Nevada Las Vegas, Las Vegas, Nevada 89154, USA}
\author{Peter Kissin}
\affiliation{Department of Physics, University of California San Diego, 9500 Gilman Drive, La Jolla, CA 92093, USA}
\author{Jacob Schalch}
\affiliation{Department of Physics, University of California San Diego, 9500 Gilman Drive, La Jolla, CA 92093, USA}
\author{Xiaomeng Cui}
\affiliation{Department of Physics, University of California, Santa Barbara, CA, 93106 USA}
\author{Sheikh Rubaiat Ul Haque}
\affiliation{Department of Physics, University of California San Diego, 9500 Gilman Drive, La Jolla, CA 92093, USA}
\author{Alex Potts}
\affiliation{Department of Physics, University of California, Santa Barbara, CA, 93106 USA}
\affiliation{Lake Shore Cryotronics, 575 McCorkle Blvd, Westerville, OH 43082, USA}
\author{Takashi Taniguchi}
\affiliation{National Institute for Materials Science, Tsukuba, Ibaraki 305-0044, Japan}
\author{Kenji Watanabe}
\affiliation{National Institute for Materials Science, Tsukuba, Ibaraki 305-0044, Japan}
\author{Richard D. Averitt}
\affiliation{Department of Physics, University of California San Diego, 9500 Gilman Drive, La Jolla, CA 92093, USA}
\author{Andrea F. Young}
\affiliation{Department of Physics, University of California, Santa Barbara, CA, 93106 USA}

\date{\today}

\begin{abstract}
The efficient modulation and control of ultrafast signals on-chip is of central importance in terahertz (THz) communications and a promising route toward sub-diffraction limit THz spectroscopy. Two-dimensional (2D) materials may provide a platform for these endeavors. We explore this potential, integrating high-quality graphene p-n junctions within two types of planar transmission line circuits to modulate and emit picosecond pulses. In a coplanar stripline geometry, we demonstrate electrical modulation of THz signal transmission by 95\%. In a Goubau waveguide geometry, we achieve complete gate-tunable control over THz emission from a photoexcited graphene junction. These studies inform the development of on-chip signal manipulation and highlight prospects for 2D materials in THz applications.     
\end{abstract}

\maketitle 

Emerging chip-scale technologies operating in the terahertz (THz) band \cite{sengupta_terahertz_2018, sengupta_chip-scale_2019, wang_voltage-actuated_2019, wu_programmable_2019} promise compact devices for sensing\cite{sizov_thz_2010}, imaging\cite{chan_imaging_2007}, security\cite{kawase_non-destructive_2003}, and communications.\cite{yang_understanding_2012, moon_long-path_2015, federici_review_2016} 
A vital requirement for these pursuits is the design of interconnects, sources, and modulators to control and guide high frequency signals on chip. Two-dimensional (2D) materials are an appealing option owing to their ultrafast charge carrier dynamics and intrinsic compactness.  Prior work has demonstrated sensitive THz detectors\cite{vicarelli_graphene_2012,spirito_high_2014,mittendorff_ultrafast_2013, bandurin_resonant_2018,viti_black_2015, murphy_terahertz_2018, rogalski_two-dimensional_2019} and transistors with cutoff frequencies reaching 50\cite{cheng_few-layer_2014} and even 350 GHz\cite{wu_state---art_2012-1}. Modulation of free space THz radiation has also been demonstrated using graphene\cite{wei_high-performance_2019, kakenov_graphene_2018, huang_broadband_2018, sensale-rodriguez_extraordinary_2012, ju_graphene_2011, li_active_2015, zhang_tunable_2018, mittendorff_graphene-based_2017} and molybdenum disulfide (MoS$_2$)\cite{chen_ultrasensitive_2016, srivastava_mos2_2017}, achieving modulation depths of 100\% and modulation speeds as high as 110 MHz.\cite{liang_integrated_2015} Harnessing ultrafast signals on-chip may additionally enable new spectroscopic measurements for fundamental science. Confinement of THz radiation in planar transmission lines enhances spatial resolution, allowing spectroscopy of materials well beyond the diffraction limit for free space THz measurements. 
Recent examples include probes of the optical conductivity of graphene in both thermal equilibrium \cite{gallagher_quantum-critical_2019} as well as under intense optical drive\cite{mciver_light-induced_2019}, and picosecond magnetization reversal in GdFeCo.\cite{yang_ultrafast_2017}  

We investigate two on-chip spectrometer designs for THz modulation and emission using ultra clean graphene p-n junctions. In coplanar striplines (CS) and Goubau waveguides (GW) we integrate graphene van der Waals heterostructures to directly modulate the line impedance using a local finger gate. Applying a gate voltage allows electrical control of on-chip THz transmission and emission. At low temperatures, we achieve 95\% modulation depth in CS circuits, which are well explained by finite-difference time-domain simulations. Finally, introducting a pump beam to directly photoexcite the graphene junction in the GW geometry, we demonstrate gate-tunable emission from the graphene junction.

\begin{figure}
\includegraphics[width=3.3 in]{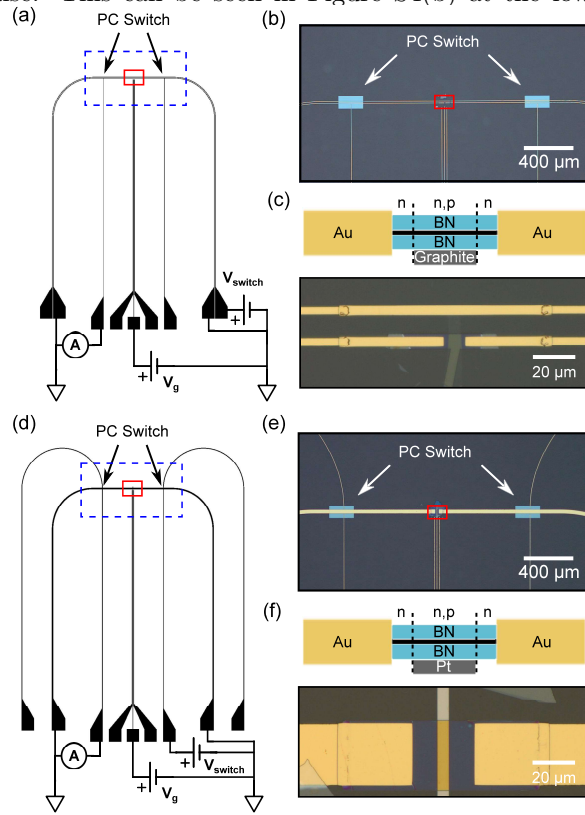}
\caption{\label{fig1} Two on-chip THz circuits with integrated graphene p-n junctions.
(a) Circuit diagram of the coplanar stripline (CS) geometry. The blue (dashed) and red (solid) boxes indicate the positions of the optical images in panels (b) and (c), respectively. The photoconductive (PC) switches are labeled with black arrows. 
(b) Optical image showing the locations of the PC switches used for both generation and detection of on-chip THz transients. The red box shows the location of the optical image in (c). 
(c) Cross-sectional model representation (above) and optical image (below) of the integrated graphene p-n junction.
(d) Circuit diagram of the Goubau waveguide (GW) geometry. The blue (dashed) and red (solid) boxes correspond to the position of the optical images in panels (e) and (f), respectively. 
(e) Optical image of the PC switches in the GW circuit.
(f) Cross-sectional model representation (above) and optical image (below) of the integrated graphene p-n junction. 
}
\end{figure}

Our high frequency circuits contain a combination of photoconductive (PC) switches and graphene p-n junctions for the purpose of generation, modulation, and detection of on-chip THz transients (Figure \ref{fig1}). Our devices are fabricated starting from silicon-on-sapphire (SOI) wafers, which are then dosed at $1\times 10^{15}$ ions/cm$^2$ with 100 keV oxygen ions. A subtractive process removes the silicon everywhere except where PC switches are desired, and waveguide circuits are subsequently patterned by standard photolithography and evaporation of titanium (10 nm) and gold (100 nm). The CS circuit (Figure \ref{fig1}a-c) consists of two 5 $\mu$m wide electrodes separated by 10 $\mu$m, resulting in a characteristic impedance of 121 $\Omega$.\cite{gevorgian_line_2001} The GW circuit (Figure \ref{fig1}d-f) consists of a single 30 $\mu$m wide conductor, resulting in a  characteristic impedance of $\approx 120$ $\Omega$.\cite{russell_broadband_2013} In both circuits we employ two PC switches (Figure \ref{fig1}b and e) with a carrier relaxation time of 560 fs determined by time-resolved THz spectroscopy (Figure S1). The right switch is photoexcited with a pulse from an amplified femtosecond laser with a 200 kHz repetition rate, 800 nm wavelength, and 25 fs pulse duration, which generates a THz transient. The left switch is then used to sample the time-domain profile of the propagating THz electric field transients using time-delayed pulses from the same laser. Figure S2 in the Supplementary Material shows a schematic of the setup and a detailed description of this measurement.

\begin{center}
\begin{figure}[h]
\includegraphics [width=6.7 in]{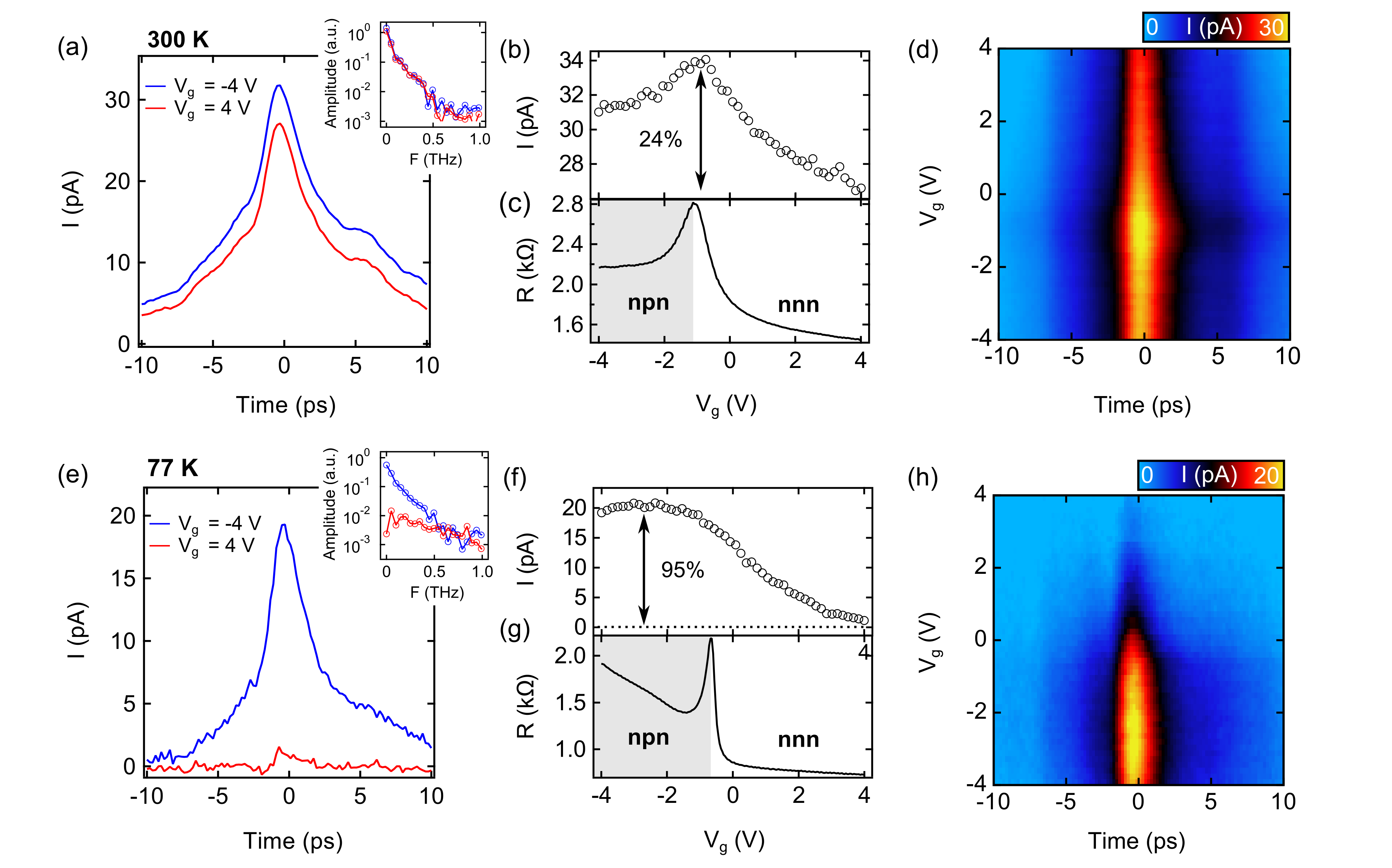}
\caption{\label{fig2} THz modulation in the CS circuit at room temperature (panels (a-d)) and 77 K (panels (e-h)). 
(a) Readout current ($I$) plotted as a function of transient time delay ($Time$) for two gate voltages ($V_g$). The inset shows the Fourier transform of the time domain scans in the main panel. 
(b) Peak current ($I$) plotted as a function of gate voltage ($V_g$).  
(c) DC resistance ($R$) as a function of $V_g$. The npn and nnn labels signify the p-n junctions created by residual doping and photodoping. 
(d) Colorplot of $I$ as a function of $V_{g}$ and $Time$ for all gate voltages explored.
(e) $I$ vs. $Time$ at 77 K. The inset shows the Fourier transform of the time domain scans in the main panel
(f) Peak current ($I$) plotted as a function of $V_g$.
(g) $R$ as a function of $V_g$ at 77 K.
(h) Colorplot of $I$ as a function of $V_{g}$ and $Time$ at 77 K.
}
\end{figure}
\end{center}

\begin{figure}
\includegraphics{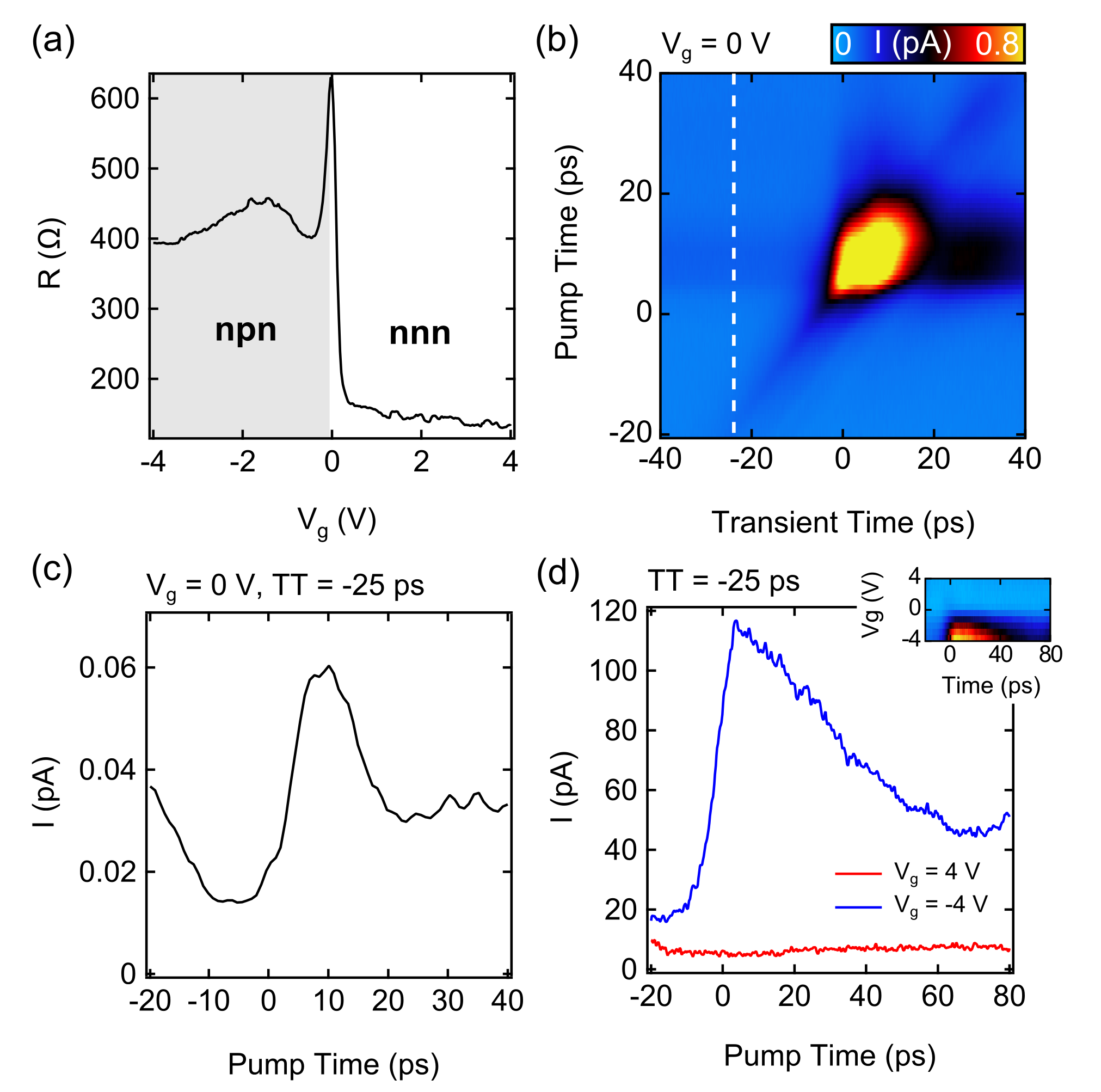}
\caption{\label{fig4} Gate tunable emission in the GW circuit at 77 K. (a) DC resistance ($R$) as a function of gate voltage ($V_g$) at 77 K for the GW graphene p-n junction. (b) Colorplot of the readout current ($I$) as a function of the pump ($Pump$ $Time$) and transient ($Transient$ $Time$ ($TT$)) time delay. The dashed line at $TT = -25$ ps corresponds to the linecuts shown in panels (c) and (d). (c) Readout current ($I$) as a function of $Pump$ $Time$ at $V_g=0$ V and $TT = -25$ ps. (d) $I$ as a function of $Pump$ $Time$ for two gate voltages, $V_g= -4, 4$ V at $Transient$ $Time = -25$ ps. The top inset shows a colorplot of $I$ vs. $V_g$ and pump time ($Time$) for all gate voltages explored. Blue corresponds to low current and yellow to high. 
}
\end{figure}

In both designs we incorporate a graphene p-n junction which directly modulates the line impedance. In the CS circuit, the ground line between the two PC switches is interrupted by a graphene junction comprised of a monolayer graphene flake sandwiched between two boron nitride (hBN) flakes with a graphite flake employed as a finger gate (Figure \ref{fig1}c).  In the GW circuit, the graphene junction forms a portion of the center conductor (Figure \ref{fig1}f). The carrier density is modulated at the center of the graphene with a platinum finger gate. In both devices, the gate is electrically isolated to reduce coupling of the transient pulses through the gate electrodes. The van der Waals heterostructures are stacked using the dry polymer method and edge contacted with chromium/palladium/gold (3 nm/15 nm/100 nm).\cite{wang_one-dimensional_2013} This junction geometry does not allow for complete control over the p-n junction response as there are ungated regions on both sides of the finger gates, as shown in the cross-sectional models in Figures \ref{fig1}(c,f). Ideally, the ungated regions are undoped. However, we observe an asymmetry in the gate dependent transport---especially pronounced at low temperatures---that points to residual n-type doping (Figure \ref{fig2}(g) and \ref{fig4}(a)). This may arise from traps at the hBN-sapphire interface as well as photodoping by the generation and detection laser pulses\cite{ju_photoinduced_2014} (Figure S3).

At room temperature, we achieve modest modulation of THz electric field transients in both geometries. Figure \ref{fig2}(a) shows the transmission time domain scans for the CS circuit at two different gate voltages. The readout current from the left PS is plotted as a function of time delay between the generation and detection laser pulses. Time zero has been arbitrarily chosen to coincide with the transient arrival at the readout switch. The transmitted transient is positive and the Fourier transform reveals frequency components up to roughly 400 GHz (inset of Figure \ref{fig2}(a)). By varying the backgate voltage, the transient amplitude is modulated. Figure \ref{fig2}(b) shows the peak amplitude as a function of gate voltage where a total signal modulation of 24\% is achieved. A colorscale plot of the readout current as a function of gate voltage and time delay is shown in Figure \ref{fig2}(d). The modulation follows the the DC characteristics (see Supplementary Materials for mobility and residual doping estimates) of the device, shown in Figure \ref{fig2}(c), with the CS transient amplitude increasing with increased channel resistance. The modulation of the picosecond transient is also directly correlated with the gate dependent transport, showing the same asymmetry between npn and nnn doping configurations. From the reflection coefficient for a series connected impedance ($\Gamma=Z/(Z+2Z_0)$), where $Z$ is the characteristic impedance of the transmission line and $Z_0$ is the graphene impedance\cite{mongia_rf_2007}, we would expect a decrease in the transmitted transient amplitude with increased $Z_0$, making the experimental results somewhat counterintuitive. 

To better understand the time-domain response of our CS circuit, we carried out finite-difference time-domain simulations\cite{CST}, shown in Figure S4. We find that when the simulated chemical potential is small and the graphene has the highest resistance, most of the transient is reflected because the ground line is effectively disconnected. Furthermore, at small chemical potentials, the amplitude of transmission inverts (changing sign) and a negative input pulse becomes and positive transmitted pulse. This can be seen in Figure S4(b) at the lowest plotted chemical potential of $\mu=0.01$ eV. The amplitude of this peak is maximized at low chemical potential (high graphene resistance), matching the observations in Figure \ref{fig2}(b,c). This regime allows for nearly complete modulation of the transmission in the CS circuit at lower temperatures where the p-n junction response is more dramatic.  

Upon cooling the circuit, we see enhanced effects of the p-n junctions. Figure \ref{fig2}(e) shows the transmission time domain scans for the CS circuit again but now at 77 K. A clear difference in the transient amplitude is discernible in the time domain and Fourier transform (inset of Figure \ref{fig2}(e)) for two different gate voltages. The peak amplitude as a function of gate voltage is shown in Figure \ref{fig2}(f) where the modulation now reaches 95\%. \onecolumngrid

\twocolumngrid A more dramatic asymmetry manifests in the gate dependent transport, with p-type carriers recording a higher resistance than n-type carriers. At $V_g=-4$ V the graphene p-n junction has the highest resistance and therefore the largest transmission amplitude. At $V_g=4$ V the graphene p-n junction has the lowest resistance and the transmission is nearly completely suppressed. A colorplot in Figure \ref{fig2}(h) shows the readout current as a function of gate voltage and time delay contrasting the peak modulation in Figure \ref{fig2}(d). 

In the GW circuit the modulation at low temperature is not as effective because the junction is better impedance matched to the waveguide (see Figure S5). However, integrating the p-n junction directly into the center conductor affords the ability to control THz emission. This is demonstrated by incorporating a third beam which directly excites the graphene junction itself. Figure S6 in the Supplementary Material shows a schematic of this measurement where three beams and two delays are used to perform an on-chip pump-probe experiment. Figure \ref{fig4}(a) shows the DC characteristics of the GW circuit junction at 77 K. We again record a clear asymmetry in hole and electron transport from the p-n junction. Figure \ref{fig4}(b) shows a 2D plot of the readout current at the left PS versus the pump and transient time ($TT$) delays, taken at $V_g=0$ V. There are three discernible features that run horizontally, vertically and diagonally through the scan which converge near the center of the plot. The highest recorded current occurs at this point when the on-chip transient pulse arrives at the graphene at the same time as the free-space pump pulse. The vertical and diagonal features correspond to the transient pulse and the pump induced change to the transient pulse, respectively. 

We also identify a horizontal feature--i.e., a feature independent of the transient pulse--which we ascribe to emission from the graphene junction itself. Picosecond pulse emission has been previously studied in 2D materials including graphene\cite{prechtel_time-resolved_2012, hunter_-chip_2015, brenneis_thz-circuits_2016}, MoS$_2$\cite{parzinger_contact_2017-1}, and bismuth selenide (Bi$_2$Se$_3$),\cite{kastl_ultrafast_2015-1}. In graphene, emission can occur as a result of photocurrents driven by an applied voltage\cite{hunter_-chip_2015} or from the photothermoelectric  effect, generated by current flow between closely spaced regions with dissimilar doping, with examples including both contact-induced\cite{hunter_-chip_2015} and gate controlled p-n junctions\cite{brenneis_thz-circuits_2016}. In our device the graphene junction is grounded on both sides, so there is no applied voltage that would give rise to emission from a fast photoconductive mechanism. Instead, we see a direct correlation between the emission and the state of the p-n junction, suggesting that the emission mechanism is a result of a photothermal voltage from the photothermoelectric effect which drives an ultrafast photocurrent accross the junction. With access to both nnn and npn junctions at various gate voltages (Fig. \ref{fig4}(a)), we can effectively turn on and off the dissimilar doping interface and thereby control emission. Figure \ref{fig4}(c) shows the emission signal well before the arrival of the transient, corresponding to the dashed white line in Fig. \ref{fig4}(b).  

At $V_g=0$ V the emission peak amplitude is weak as the center of the junction at charge neutrality, so that no sharp heterointerface is present to give rise to photothermolelectric currents. By tuning the gate voltage, however, the emission amplitude can be strongly enhanced. Fig. \ref{fig4}d shows the time domain scan of the pump signal at $V_g=\pm4$ V and $TT = -25$ ps. At $V_g=-4$ V, deep in the npn state, we record strong high frequency emission from the graphene junction. By tuning the junction into the nnn state, $V_g=4$ V, the emission is suppressed. The inset of Fig. \ref{fig4}(d) shows the complete evolution of these time domain scans with gate voltage. Various heat dissipation channels through the substrate and contacts could contribute to the slower time response we record here when compared to ultrafast photothermoelectric signals observed in other junctions.\cite{brenneis_thz-circuits_2016}

Summarizing, we investigated the potential of integrated graphene junctions in high frequency circuits to control transmission and emission of THz transients on-chip. Our results reveal that 2D materials have strong potential in high frequency applications as tunable sources and modulators. The explored geometries additionally lend themselves to spectroscopic studies of nanomaterials with dimensions below the diffraction limit at THz frequencies. 

\section{Supplementary Material}
The Supplementary Material includes the carrier relaxation times of the Si photoconductive switches, a description of the transient generation and detection measurement, details about p-n junction creation, mobilities, finite difference time domain simulations of the CS circuit, modulation characteristics of the GW circuit, and a description of the pump beam measurements. 

\begin{acknowledgments}
This work was primarily supported by the Army Research Office under MURI W911NF-16-1-0361.
J.O.I. acknowledges the support of the Netherlands Organization for Scientific Research (NWO) through the Rubicon grant, project number 680-50-1525/2474. A portion of this work was performed at the Institute for Terahertz Science and Technology (ITST) at UCSB. 
\end{acknowledgments}

\section{Data Availability}
The data that support the findings of this study are available from the corresponding author upon reasonable request.
\section{References} 
\bibliography{Integrated}

\clearpage
\onecolumngrid
\section{Supplementary materials: On-chip terahertz modulation and emission with integrated graphene junctions}

\setcounter{equation}{0}
\setcounter{figure}{0}
\setcounter{table}{0}
\setcounter{page}{1}
\setcounter{section}{0}
\renewcommand{\thefigure}{S\arabic{figure}}
\renewcommand{\thesubsection}{S\arabic{subsection}}
\renewcommand{\theequation}{S\arabic{equation}}
\renewcommand{\thetable}{S\arabic{table}}
\renewcommand{\thepage}{S-\arabic{page}}

\subsection{Relaxation time in radiation damaged silicon}
\begin{figure}[ht]
\includegraphics{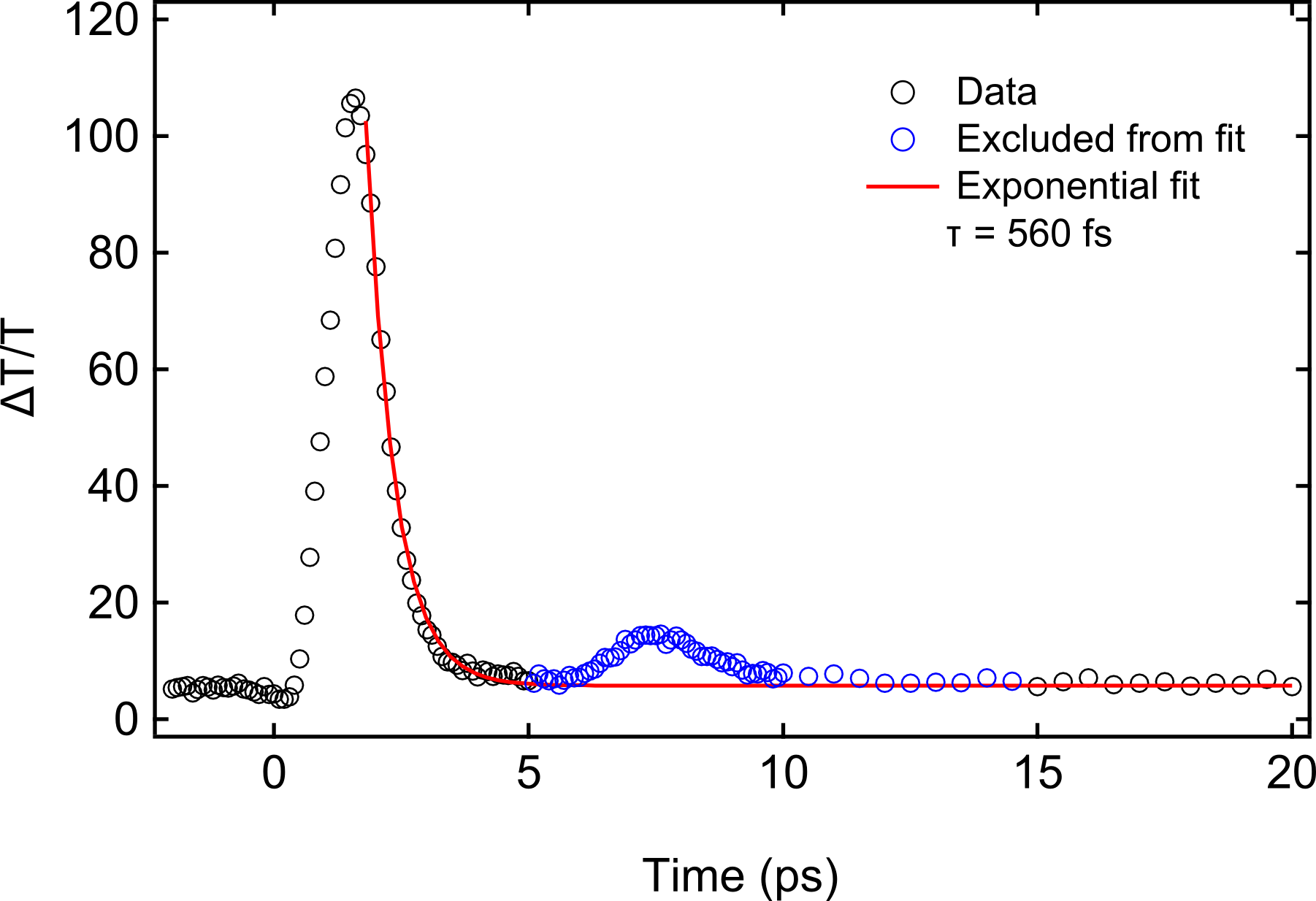}
\caption{\label{fig_S3} Relaxation time for the radiation damaged silicon. The data in blue (second smaller peak) a spurious effect as described in the text, has been excluded from the single exponential fit. 
}
\end{figure}

In order to determine the carrier relaxation time of our radiation damaged silicon, we performed time-resolved THz spectroscopy. Figure \ref{fig_S3} shows the THz transmission through a sample of radiation damaged silicon on sapphire as a function of the time delay of an optical pulse from an amplified femtosecond laser used to photoexcite the sample. The main peak tracks the relaxation dynamics of photoexcited carriers in the radiation damaged silicon. The spurious second peak is due to a reflection of the optical pulse from back of the sapphire substrate, which re-excites the radiation damaged silicon. We extract a relaxation time of 560 fs from a single exponential fit to the first peak, excluding the second peak. 

\subsection{Transient generation and detection measurements}\label{sec1}
\begin{figure}[ht]
\includegraphics [width=5 in]{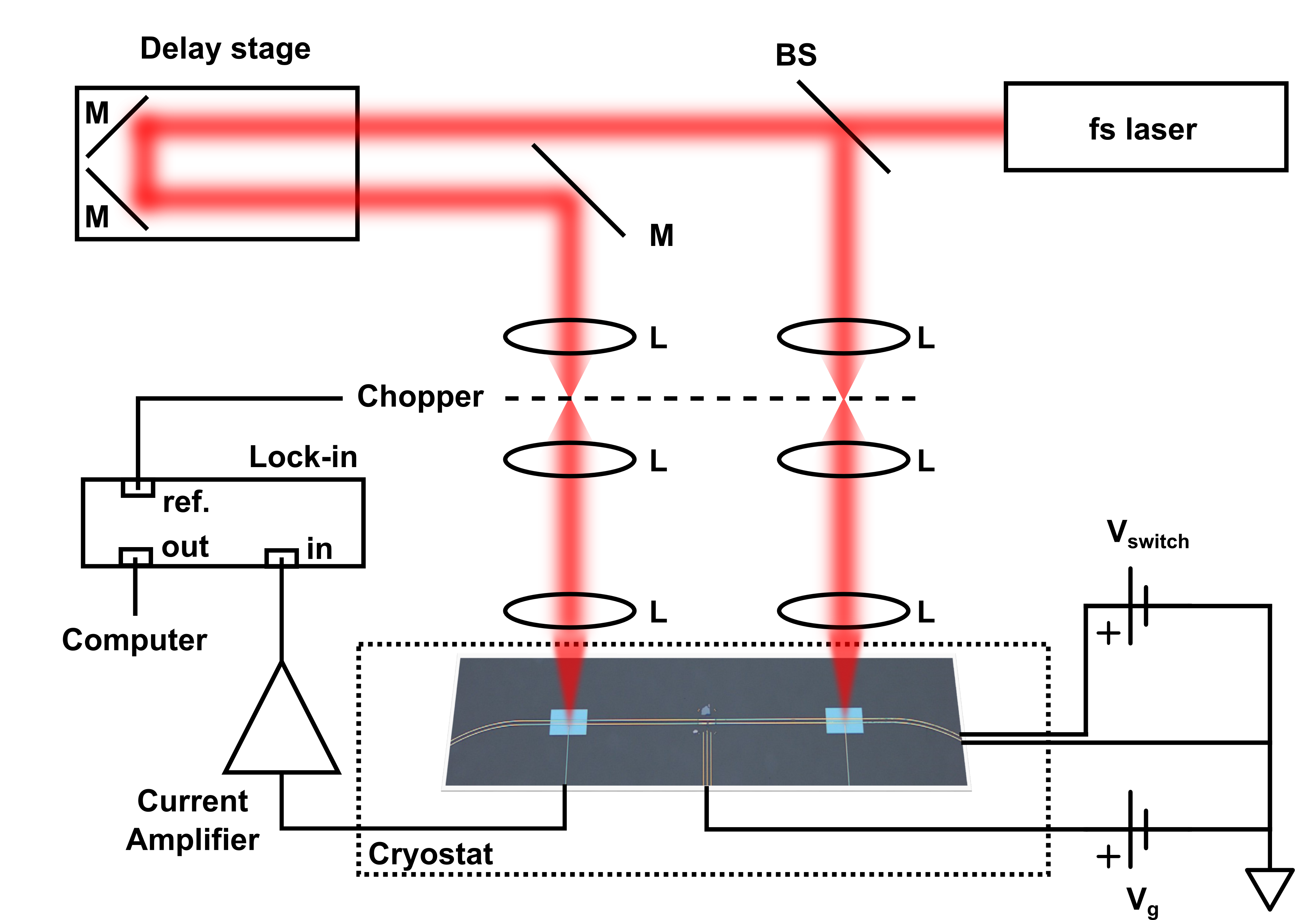}
\caption{\label{fig_S1} Schematic of the THz transient generation and detection measurement corresponding to data presented in the main text in Figure 2. 
}
\end{figure}

The on-chip transient generation and detection is performed using a time delayed femtosecond laser. Figure \ref{fig_S1} shows a schematic representation of the measurement method. A beamsplitter (BS) splits the beam into two paths. One is directed through a mechanical chopper and focused on the right photoconductive switch (PS) to generate the transient. The second is sent to a time delay stage, directed through the chopper and focused on the left photoconductive switch to sample the transient profile. The two beams are modulated at two different frequencies and the signal is detected with a lock-in amplifier at the sum or difference of the two frequencies. The current from the left photoconductive switch is amplified and then sent to the lock-in. The transmission line is biased and a voltage is applied to the gate using DC voltage supplies. The cryostat, indicated with the dashed box, is evacuated and kept at room temperature or cooled for low temperature measurements.  

\subsection{p-n junction creation by photodoping} 
\begin{figure}
\includegraphics{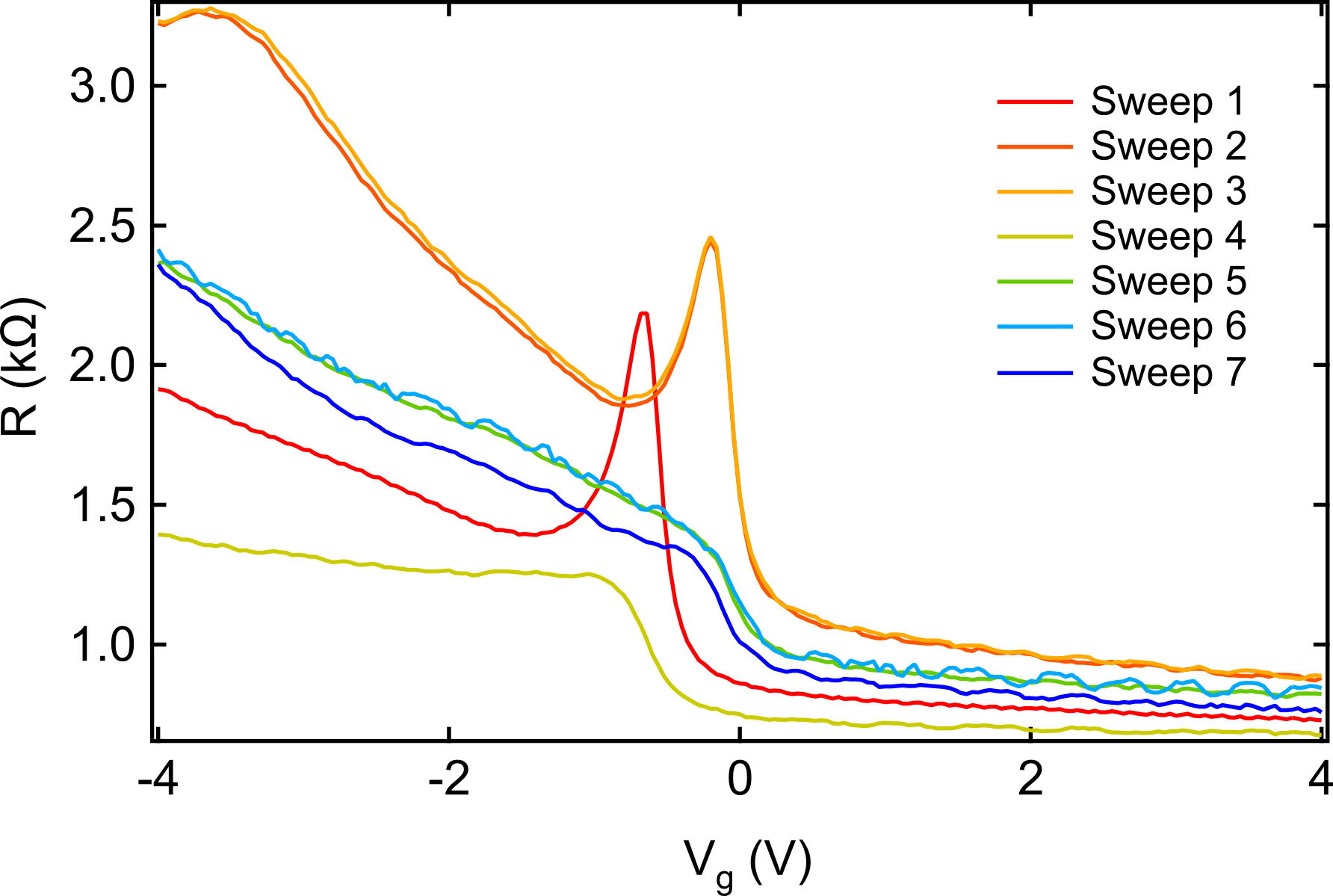}
\caption{\label{fig_S4} Temporal change in gate sweeps for the CS graphene junction due to varying photodoping. Resistance $R$ plotted as a function of gate voltage $V_g$ for several sweeps taken at various times.  
}
\end{figure}

Our p-n junctions form as a result of residual doping and photodoping from scattered light coming from the laser pulses used to generate and detect THz transients. While advantageous to achieve high modulation depths and tunable emission, the p-n junction response changes over time with variations in doping of the graphene channel due to trap states in the boron nitride. Figure \ref{fig_S4} shows several gate sweeps of the CS graphene junction over various time intervals. Sweep 1 was taken just before the 2D scan shown in Figure 2(h). Subsequent scans show qualitative changes as the trap states fill around the gated and ungated regions of the graphene junction over time. It is therefore important that the DC and high frequency measurements are taken with minimal time delay in between.   

\subsection{Mobility and residual doping estimates} 
From the DC characteristics we estimate the mobilities and residual charge doping. The maximum electron mobility estimated from the data in Figure 2(C) is $\mu = 1/neR \approx 130,000$ cm$^2$/Vs, where $n=C_gV_g/e$ is the charge density. $C_g=\epsilon_{BN}\epsilon_0/t_{BN}$ is the gate capacitance per area with $t_{BN}\approx40$ nm, the bottom BN thickness and $\epsilon_{BN}=3$, the relative permitivity of BN. The offset of charge neutrality from zero back gate voltage amounts to a residual carrier doping of $n \approx 4.5\times 10^{11}$ cm$^{-2}$.
From the DC characteristics for the low temperature data, Figure 2(g), we estimate a maximum electron mobility of $\mu \approx 190,000$ cm$^2$/Vs and a smaller residual carrier doping of $n \approx 2.7\times 10^{11}$ cm$^-2$. 
For the GW device at low temperature, Figure 3(a), We estimate a maximum electron mobility of $\mu \approx 530,000$ cm$^2$/Vs and  record negligible residual doping at charge neutrality. For this device the bottom BN thickness is $\approx 50$ nm. 

\subsection{Finite difference time domain (FDTD) simulations for the coplanar stripline device} 
\begin{figure}
\includegraphics [width=6.7 in]{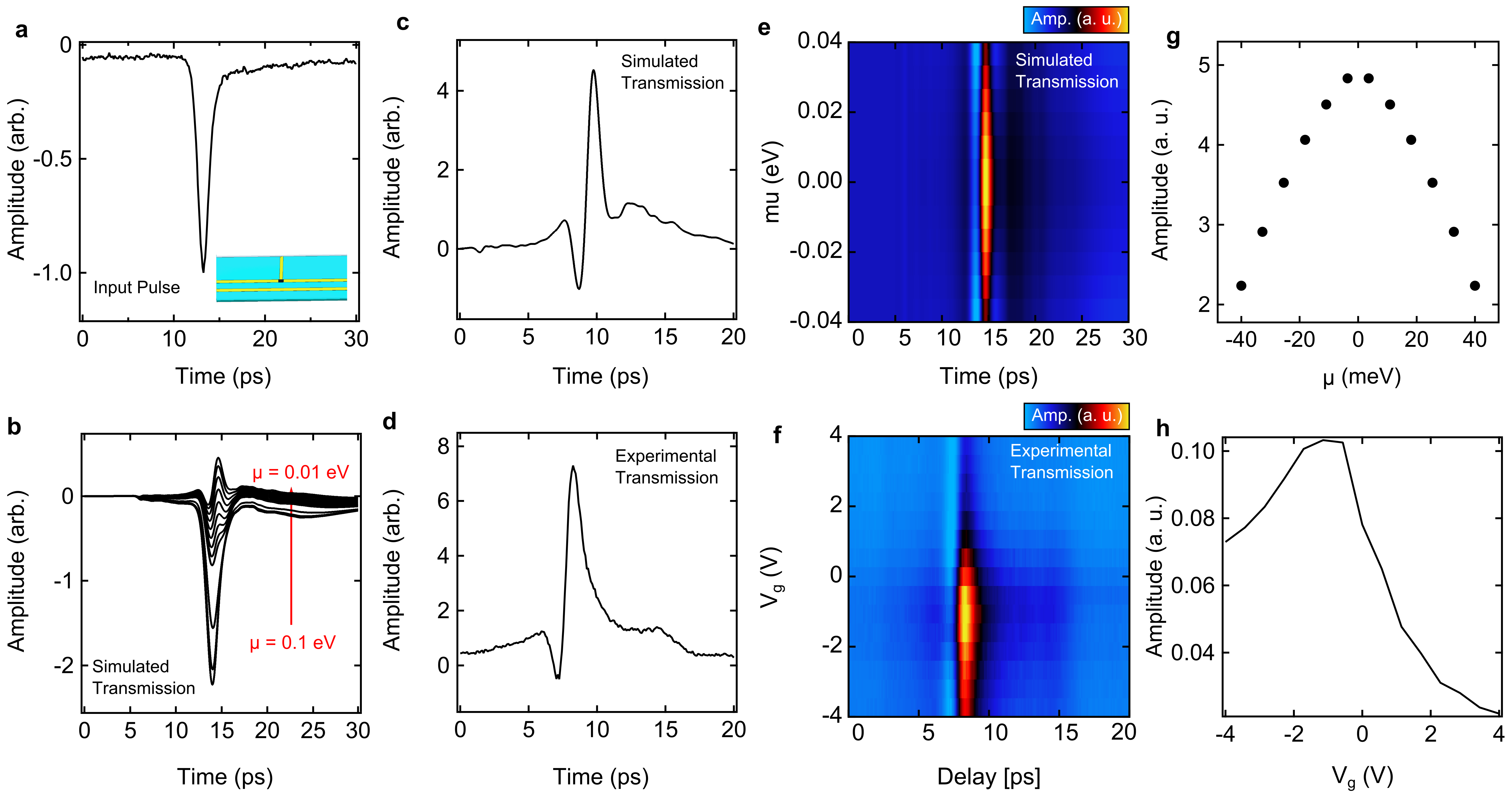}
\caption{\label{fig_S5} CST simulations showing peak inversion at low chemical potentials
(a) Input pulse for CST simulations provided by a transient measured on a CS without a graphene junction. The inset shows an image of the CST modeled CS circuit with a graphene junction on the ground line.  
(b) Simulated transmission of the transient as a function of chemical potential of the graphene. A peak emerges at low chemical potentials. 
(c) Simulated transmission at a chemical potential of 0.02 eV. 
(d) Experimental transmission in a CS circuit with a graphene junction measured at 5 K. 
(e) Evolution of the simulated transmission as a function of low chemical potential. The negative values have been mirrored from the positive chemical potential calculations.
(f) Experimental transmission as a function of gate voltage on the graphene. The red/yellow colors indicate higher amplitude and the blue lower. 
(g) Simulated transient amplitude as a function of chemical potential of the graphene. 
(h) Peak current for a CS circuit device taken from the data in panel (f). 
}
\end{figure}

We modeled the coplanar stripline circuit with a graphene junction on the ground line in CST. The inset of Figure \ref{fig_S5}(a) shows a 3D model of the structure. The wave is launched by a discrete waveguide port inside the CS at the right side and detected by a voltage probe outside of the CS at the left side. The graphene is modeled as a 2D impedance surface. An experimentally measured transient on a CS without a graphene junction was used as the input pulse for our simulations, Figure \ref{fig_S5}(a). This amounts to an attenuated version of a real input pulse but has the qualitative features necessary to investigate the curious gate modulated transmission amplitude mentioned in the text for the CS circuit. The simulated transmission is shown in Figure \ref{fig_S5}(b) as a function of chemical potential of the graphene. At higher chemical potential (0.1 eV) the transmitted transients have the same polarity as the input pulse and the shape is qualitatively the same. At lower chemical potentials the dip diminishes and a peak forms. This peak corresponds to the measured transient shown in Fig. 2(a). The exceedingly small chemical potential at which the crossover occurs is a result of the contact resistance of the graphene junction. The relatively large contact resistance, not included in the simulation, effectively puts the junction in this low chemical potential regime. Indeed, the qualitative shape of simulated transient (Figure \ref{fig_S5}(c)) as a function of low chemical potential (Figure \ref{fig_S5}(e)) agrees well with experimental measurements of another CS device (Figure \ref{fig_S5}(d)) as a function of gate voltage (Figure \ref{fig_S5}(f)) measured at our lowest temperature of 5 K. We note that the effects of the photodoping which leads to the p-n junction response has not been included in the simulations. The chemical potential as a function of gate voltage (Figure \ref{fig_S6}(g)) therefore does not reproduce the asymmetry in the experimental peak amplitude as a function of gate observed in Figure \ref{fig_S5}(h). However, the qualitative response is captured where the peak amplitude is maximized at low chemical potentials (high graphene resistance).

\subsection{Room temperature and low temperature modulation in the GW circuit} 
\begin{figure}
\includegraphics [width=6.7 in]{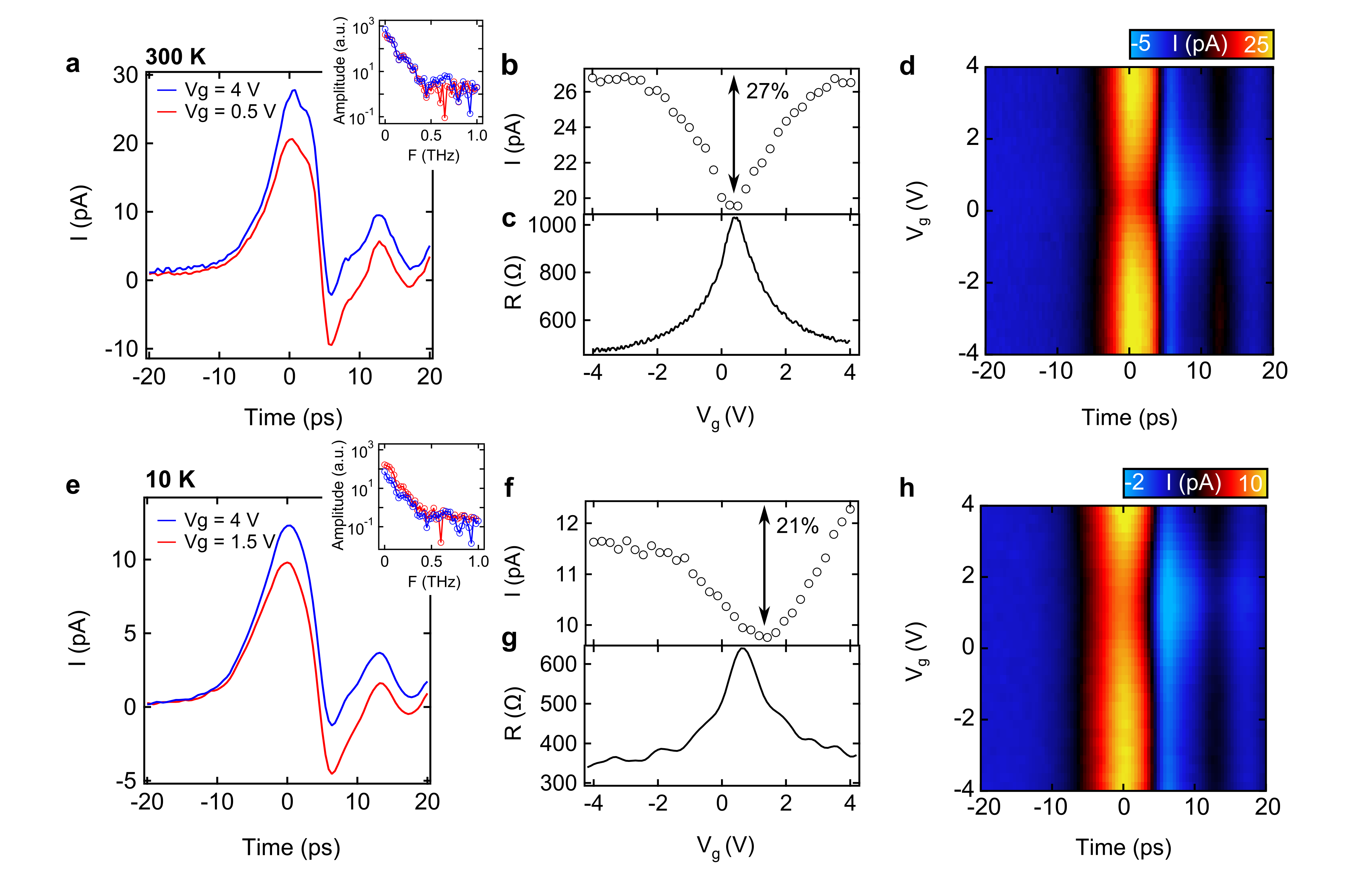}
\caption{\label{fig_S6} THz modulation in the GW circuit at room temperature (panels (a-d)) and 10 K (panels (e-h)). 
(a) Readout current ($I$) plotted as a function of transient time delay ($Time$) for two gate voltages ($V_g$). The inset shows the Fourier transform of the time domain scans in the main panel. 
(b) Peak current ($I$) plotted as a function of gate voltage ($V_g$). The transient modulation is 27\% at room temperature.  
(c) DC resistance ($R$) as a function of $V_g$.
(d) Colorplot of $I$ as a function of $V_{g}$ and $Time$ for all gate voltages explored.
(e) $I$ vs. $Time$ at 10 K. The inset shows the Fourier transform of the time domain scans in the main panel
(f) Peak current ($I$) plotted as a function of $V_g$. The transient modulation is 21\% at 10 K.  
(g) $R$ as a function of $V_g$ at 10 K.
(h) Colorplot of $I$ as a function of $V_{g}$ and $Time$ at 10 K.
}
\end{figure}

As mentioned in the main text, the modulation is less effective in the GW circuit mainly due to the more optimal impedance matching between the graphene junction and the waveguide. Figure \ref{fig_S6} shows the room temperature (Figure \ref{fig_S6}(a-d)) and low temperature (Figure \ref{fig_S6}(e-h)) modulation characteristics of the GW circuit. The peak modulation (Figures \ref{fig_S6}(b,f)) is opposite the DC characteristics (Figures \ref{fig_S6}(c,g), respectively). This is expected for a series connected impedance (the reflection coefficient increases with increased graphene impedance $\Gamma=Z/(Z+2Z_0)$) and evidences the better impedance matching in this circuit when compared with the gate dependence of the DC resistance for the CS circuit outlined in the main text and detailed with CST simulations. The low temperature modulation at 10 K is slightly reduced (21\%) as the DC characteristics show less change in resistance between charge neutrality and higher gate voltages (Figure \ref{fig_S6}(g)).   

\subsection{Transient generation and detection measurements with a pump beam}
\begin{figure}
\includegraphics [width=5 in]{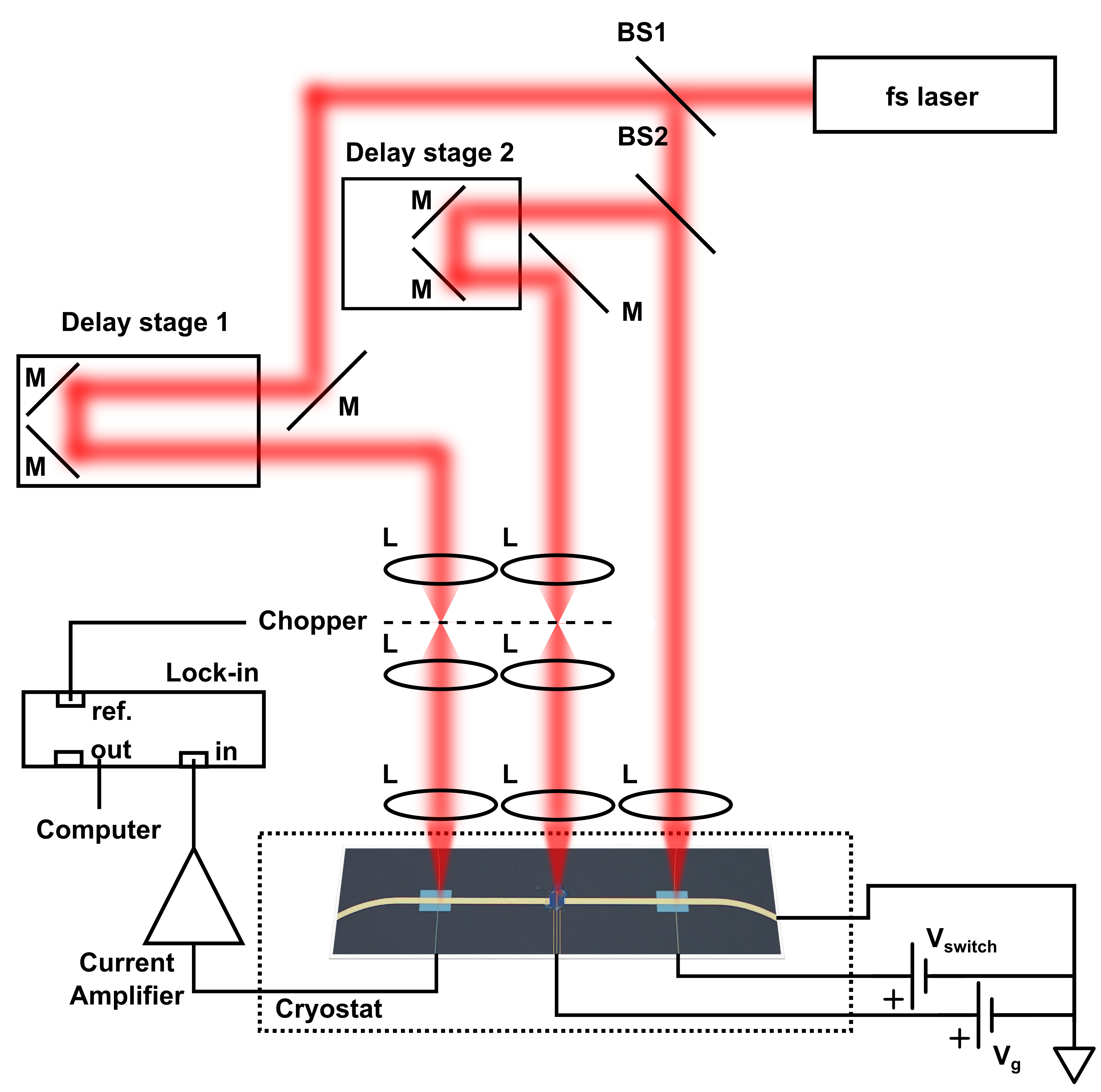}
\caption{\label{fig_S2} Schematic of the THz transient generation and detection measurement with a pump beam to photoexcite the graphene junction. This setup corresponds to data presented in the main text in Figure 3.
}
\end{figure}

In the pump beam measurements, see Figure \ref{fig_S2}, the generation and detection of the transient is carried out as in Section \ref{sec1} but the laser is split a second time with another beamsplitter (BS2) and sent to a second delay stage (Delay stage 2) to control a third beam which is focused directly on the graphene junction. The current is read out from the left photoconductive switch and voltage supplies again control the bias on the generation switch and graphene gate.

\end{document}